# A neural network model for timing control with reinforcement

Jing Wang[1], Yousuf El-Jayyousi[1], Ilker Ozden[1]

Department of Biomedical, Biological, and Chemical Engineering, University of Missouri, Columbia, MO, United States

**\* Correspondence:** ozdeni@missouri.edu

**ABSTRACT**

How do humans and animals perform trial-and-error learning when the space of possibilities is infinite? In a previous study (Wang et al., 2020), we used an interval timing production task and discovered an updating strategy in which the agent adjusted the behavioral and neuronal noise for exploration. In the experiment, human subjects proactively generated a series of timed motor outputs. Positive or negative feedback was provided after each response based on the timing accuracy. We found that the sequential motor timing varied at two temporal scales: long-term correlation around the target interval due to memory drifts and short-term adjustments of timing variability according to feedback. We have previously described these features of timing variability with an augmented Gaussian process, termed reward sensitive Gaussian process (RSGP). In a nutshell, the temporal covariance of the timing variable was updated based on the feedback history to recreate the two behavioral characteristics mentioned above. However, the RSGP was mainly descriptive and lacked a neurobiological basis of how the reward feedback can be used by a neural circuit to adjust motor variability. Here we provide a mechanistic model and simulate the process by borrowing the architecture of recurrent neural networks. While recurrent connection provided the long-term serial correlation in motor timing, to facilitate reward-driven short-term variations, we introduced reward-dependent variability in the network connectivity, inspired by the stochastic nature of synaptic transmission in the brain. Our model was able to recursively generate an output sequence incorporating the internal variability and external reinforcement in a Bayesian framework. We show that the model can learn the key features of human behavior. Unlike other neural network models that search for unique network connectivity for the best match between the model prediction and observation, this model can estimate the uncertainty associated with each outcome and thus did a better job in teasing apart adjustable task-relevant variability from unexplained variability. The proposed artificial neural network model parallels the mechanisms of information processing in neural systems and can extend the framework of brain-inspired reinforcement learning in continuous state control.

1. INTRODUCTION

Reinforcement learning (RL) is situated at the growing intersection between machine learning and learning rules in biological systems. RL theory has been highly influential in explaining the rules an agent applies in interacting with the environment. In the conventional RL paradigm, a discrete set of choices are presented. The agent learns the values associated with individual options and reactions in order to optimize control policies and thus maximize the reward outcome (Sutton and Barto, 1998). However, we often face choices of continuous nature, such as deciding the serving angle of a tennis ball. In such cases, it is impossible to learn the entire span of value functions with a finite number of attempts (Dhawale et al., 2017; van Hasselt, 2012; van Hasselt and Wiering, 2007). A variety of learning algorithms have been proposed to address this problem (Benbrahim and Franklin, 1997; Dhawale et al., 2017; Gullapalli, 1990; van Hasselt, 2012; Wu et al., 2014). These algorithms mainly focus on constructing models for stochastic exploration of the reward landscape in the space of model parameters.

Our previous study (Wang et al., 2020) described an updating strategy in which the agent can actively explore the space of choices by adjusting their behavioral noise (motor variability). In the experiment, test subjects proactively generated a series of internally timed motor outputs by adjusting their timing accuracy trial-by-trial in response to positive or negative feedback. We found that the time series consisting of humans' sequential motor responses varied at a slow and a fast temporal scale: the long-term serial correlations due to memory drifts and a short-term adjustment of noise due to feedback. Subsequent experiments verified the causal influence of binary feedback on humans' motor variability. We previously described the observations with an augmented Gaussian process, termed reward sensitive Gaussian process (RSGP). In a nutshell, the covariance function was updated based on the feedback intrinsically recreating the two behavioral characteristics mentioned above. However, the RSGP, like previous learning algorithms mentioned above, lacked a neurobiological basis of how reward is used by a neural circuit to adjust motor variability. The goal of this study is to incorporate reinforcement learning into the time series modeling and explain our observations in a biologically plausible manner.

Accordingly, here we present a mechanistic view of the observed behavior by borrowing the architecture of recurrent units in recurrent neural networks (RNN). We modified the unit architecture in order to incorporate the reinforcement. In the recurrent unit, we introduced adjustable noise to the connectivity and linked the noise level to the reward feedback. With this configuration, we are able to estimate the uncertainty of the prediction and evaluate the prediction according to its uncertainty. We call our model a weight varying autoregressive NN (vARNN). Unlike standard machine learning models that search for unique connectivity providing the most accurate prediction, our model provides an estimate of both the mean and the uncertainty associated with each sample, which leads to the evaluation of the whole series using a Bayesian framework. Our model was able to recursively generate a sequence of timing outputs by integrating an external reinforcement signal with its internal variability.

In constructing our model, we aimed at an interpretable (biologically realistic, and mechanistic) and simple architecture. Below, we show that (1) the time series output of the vARNN can capture both temporal dependencies and reward sensitivity of the human behavior; (2) this is a generative model that can be learned and parameters can be reconstructed; and (3) by considering the uncertainty associated with the model prediction, the vARNN overall did a better job in teasing apart explainable/adjustable noise from other noise than the other models did. We will demonstrate the last point by comparing the performance of our model to two already existing generic neural network models used for modeling time series. These alternative network models, namely autoregressive neural network (ARNN) and the gated recurrent unit (GRU) model are discussed in detail below.

## 2. RELATED WORK AND OUR CONTRIBUTION

Recurrent neural networks (RNNs) are designed to handle the sequential dependence in data with temporal structure. RNN architectures with recurrent units such as Long Short Term Memory (LSTM) cells are widely used in value forecasting and language processing. They are powerful tools for making predictions based on historical data. In real life, the evolution of a time series is always influenced by multiple sources, and different sources can widely vary in how they affect the outcome. In neural network models, we are hoping that any input-output relationship can be learned with a sufficiently complex network structure and a large amount of training data. Here, we present a case and argue that if the aim is an interpretable and mechanistic model, we should take into account the nature of the information before applying any neural network model.

In modeling the time series dynamics with long-term dependencies as in our motor timing task, a vanilla ARNN has been proposed as an efficient method (Triebe et al., 2019). However, ARNN assumes a deterministic mapping from previous values (inputs) to the predicted value (output) (Triebe et al., 2019). The gated recurrent unit (GRU) model, a concise version of LSTM, is another alternative for explaining long-term temporal dependencies in the time series data. GRU can accommodate the reward feedback as part of the inputs, however, it was not instructed on how the past affects future observations. The error between the model prediction and actual observation was often attributed to the system noise that was identical among all samples and independent of the underlying process. In contrast to the ARNN and GRU, a weight varying autoregressive NN (vARNN) models the distribution of network parameters and uses feedback to adjust the weights of respective inputs. This is in line with the Bayesian net which takes into account the stochastic nature of the observations in updating the distribution of the network weights.

In vARNN, we intended to bridge the gap between the learning driven by scalar feedback (amount of reward or binary reward) and learning driven by gradients (regressive process). Even though we did not use complex structures such as multiple layers in deep networks, this model was able to learn the key features of human behavior. It can also be readily adapted to other processes that take scalar feedback into consideration. Moreover, this model extends the framework of reinforcement learning to continuous state systems.

## 3. MATERIALS AND METHODS

### 3.1. Human behavior experiment

Details of the experiment design and data collection were presented in a previous report (Wang et al., 2020). In brief, the human subjects were instructed to make delayed saccadic eye movements or button presses in response to an instruction signal (Cue-Set-Go, CSG task. Figure 1a). The duration of the delay (production time $t_p$) was compared to a target duration ($t_t$), which was unbeknown to the subjects and was randomly sampled from a normal distribution $\sim \mathcal{N}(800ms, 80ms)$ between behavioral sessions. The relative error (*e*) in production time was defined as $e = (t_p - t_t)/t_t$. The subjects received positive feedback ($r = 1$) if *e* was within an acceptance window and negative feedback ($r = 0$) if it was outside the window (Figure 1c). The acceptance window was determined based on each subject's performance so that approximately half of the trials led to positive feedback. In each subject, the error distribution was centered around zero and the responses were sensitive to the feedback (Figure 1b). For each experimental session, we obtained two series consisting of error (e) and reward outcome (r) in the trial sequence as indicated by $e_n$ and $r_n$ (Figure 2a).

### 3.2. Auto-regressive neural network model (ARNN) model

In the classic recurrent neural network model, the input is integrated with the internal state of the recurrent unit to produce an output. Noise can be added to the output. We conceptualized the interval generating process in the recurrent unit as follows. The recurrent unit receives an update as feedback at each trial and combines it with the past history stored in the recurrent memory. The integration between univariate input and multivariate cell memory can be expressed as

A model of order p can be expressed as
$$e_n = h(W^e \cdot e_{n-1} + W^{h^\top} h_{n-1} + b) + \epsilon_n$$
in $e_{n-1}$ is the input, $\vec{h}_{n-1} = [h_{i-1}, , h_{i-p}]$ is the cell memory. $h(.)$ is the activation function. $b$ is the bias term and $\epsilon_n$ is the independent noise sampled from a normal distribution. The weights $[W^e, W^h] = \{w_i\}$ indicate the leverage of the current trial in the past $i^{th}$ trial. The actual dimensionality of the memory input (*p*) might vary and thus be determined for individual time series. Common practice is that a set of parameters is assumed to be constant throughout the time course and optimized with respect to the ensemble training data.

### 3.3. vARNN model

The weight varying autoregressive model (vAR) requires the recurrent connection to be adjusted on the fly for every sample (Figure 3a). This is in a way similar to the stochastic nature of synaptic transmission in the brain, which is proposed to be the substrate for motor variability and stochastic exploration during motor learning (Gershman and Ölveczky, 2020; Llera-Montero et al., 2019; Seung, 2003). Therefore, unlike the classic AR assuming constant weights, we introduced additional information provided by the feedback ($r_{n-i}$) for updating the weights and introducing variability. Even though negative feedback ($r = 0$) could not inform us about the size of the error (whether the interval is too long or too short), our data showed that subjects still proactively changed their motor noise. Here, we propose that the feedback is used to tune the variance, as opposed to the strength, of the associated connection $w_i$. To keep the relationship between feedback and weights simple, we assumed that the weight variance is modulated by the feedback (*r*) in a linear fashion bounded in $[\sigma_+^2, \sigma_-^2]$, and this relation is applied ubiquitously for all weights:
$$w_i(r) \sim w_i \cdot \mathcal{N}(1, \sigma_-^2(1-r) + \sigma_+^2 r)$$
$$\epsilon_n \sim \mathcal{N}(0, \sigma_0^2)$$
$$h(x) = ReL(x+1)$$
As a special case, the vAR model reduces to the classical AR model if $\sigma_+^2$ and $\sigma_-^2$ are set to zero. The activation is rectified linear (ReL) function that ensures that the NN generates a realistic output ( i.e. $t_p >$ 0 and *e* > -1). $\epsilon_n$ is the remaining unexplained (or task irrelevant) motor noise that was sampled from a normal distribution of zero mean and $\sigma_0$ standard deviation. Though more complex hidden structures can be included in the vARNN to achieve better fitting accuracy (with the danger of overfitting), we will evaluate the model with other alternatives of similar complexity.

### 3.4. Gated recurrent unit (GRU) model

Due to vanishing or exploding gradient problems, a sequential model can not keep track of arbitrary long-term dependencies. In the ARNN, one can alleviate the issue of vanishing gradients by setting a cut-off for the autoregressive order $p$. On the other hand, the RNNs with LSTM units solve the gradient issue by controlling the sequential dependencies and the updates of internal states. The gated recurrent unit (GRU) is a simplified LSTM unit that merges the cell state with the hidden states. In our implementation, the GRU takes both the value and reward outcome as inputs.
The computation within the GRU is detailed as follows

Reset gate: $r_n = \sigma(W_r[X_n; h_{n-1}] + b_r)$
Forget gate: $f_n = \sigma(W_f[X_n; h_{n-1}] + b_f)$
Candidate for the hidden state: $\tilde{h}_n = \tanh(W_h[X_n; r_n \odot h_{n-1}] + b_h)$
Update of the hidden state: $h_n = (1 - f_n) \odot h_{n-1} + f_n \odot \tilde{h}_n$

$\odot$ symbolizes element-wise multiplication and $\sigma$ indicates sigmoid function. The GRU used h hidden units to track the hidden state. We have tested networks with various sizes of hidden units, and have shown $h = 32$ in the result as it has reached asymptotic performance. The input $X_n$ at n$^{\text{th}}$ trial was from the previous trial's $e_{n-1}$ and $r_{n-1}$, therefore the input has two dimensions. The parameters to be learned are the weight matrix $W_{r,f,h} \in \mathbb{R}^{h \times (d+h)}$ and the bias $b_{r,f,h} \in \mathbb{R}^h$.

### 3.5. Time series simulation and model learning

For ARNN and vARNN, we simulated the time series using designated sets of parameters $\theta = \{\omega_i, \sigma_+^2/\sigma_-^2\}$, where $i = 1, ...p$. To verify that the model parameters can be learned and the optimization procedure is robust, we trained the model with simulated data generated from various assigned parameters and compared the optimized parameters with the assigned ones. Similarly in GRU, we simulated the time series. However, the model parameters do not have interpretable correspondence like the ones (weight and variance ratio) in ARNN and vARNN.

All models made one step forward forecasts ($\hat{e}_n$) based on their corresponding graph in Figure 3. In all models, we assumed an additive noise ($\epsilon_n$) at the last step was irreducible and was sampled from the independent identical distribution (i.i.d.). The residual error, defined as the difference between the estimated mean $\hat{e}_n$ and actual output, was what we aimed to minimize on the training data with respect to the set of chosen parameters. In real practice, the loss function (*Loss*) consists of the residual error (*Res*), and a regularization (*P*) to penalize the model complexity. The regularization prevents overfitting and helps strike a balance between the accuracy and generalizability of the model. It has two parts, the first part aims to relax the constraint of knowing the exact AR order for each time series. The models could have included inputs from far back in time, and this term ensures setting small weights to zeros. The regularization function achieves this by having a large gradient when the weight is close to zero and a vanishing gradient when it is far away from zero. The second term keeps the ratio of weight variance in a reasonable range.

$$Loss(\theta) = Res + P$$

$$P(\theta) = \lambda_1 \left[\frac{1}{p}\sum_{i=1}^{p} \frac{2}{1 + \exp(-c_1|\omega_i|^{1/c_2})} - 1\right] + \lambda_2[\sigma_-^2/\sigma_+^2 + \sigma_+^2/\sigma_-^2]$$

In which $\lambda_1$ and $\lambda_2$ are the regularization coefficients. It relates to the data noise and we use $\lambda_1, \lambda_2 = 0.01\sigma^2(e_n)$. $c_1$ and $c_2$ are the regularization curve parameters. With AR coefficients in the range [0, 1], $c_i, c_2 = 3$ works well for the data (Triebe et al., 2019).

$$Res(\theta)^{\text{wMSE}} \equiv \frac{1}{N}\sum_{n=1}^{N} g_n[\hat{e}_n(\theta) - e_n]^2 = \frac{1}{N}\sum_{n=1}^{N} \frac{1}{\sigma_n^2}[\hat{e}_n(\theta) - e_n]^2$$

In a deterministic NN model, the optimization objective is conventionally defined by the average of all prediction errors, so called mean squared error (*MSE*). It assumes equal contribution from each data sample. In our case, vAR directly provides a non-trivial output variance which varies trial-by-trial. Aitken showed that the best linear estimator is when a weighted sum of residuals is minimized, and the weight is proportional to the reciprocal of the individual variances (Aitken, 1936). The intuition is similar to Bayesian inference in that the error from an unreliable observation (large variance) should be weighed less than the error of more reliable ones. We therefore used this weighted MSE (*wMSE*) for computing the objective function. The error weight for each trial sample is indicated by $g_n$, not to be confused with the connection weights ($\omega_i$) of the NN. A set of parameters were obtained for each simulation/experimental time series by minimizing the loss by combining the *wMSE* of residual and the penalty.

As the null estimation for the weighted residual function, we computed it with random weights:

$$Res(\theta)^{\text{rwMES}} = \sum_{n=1}^{N} g_n[\hat{e}_n(\theta) - e_n]^2 / \sum_{n=1}^{N} g_n$$

In which $g_n$ is the random number sampled uniformly from a positive value set, e.g.[0, 1]. As a special case, when all $g_n$ are equal, *wMSE* reduces to the *MSE*.

All three models were trained on segments of the time series. For ARNN and vARNN, the dimensionality ($p$) of the cell state variable might vary between individual sessions, however, we initialized it to be p = 20 as inferred from the results of partial correlation analysis of the data. For GRU, a sequence of inputs was provided and the corresponding final outputs were evaluated. In both the simulated and experimental data, models were trained on 80% of data sampled randomly without replacement, and the rest 20% were used for validation. We used a stochastic gradient descent optimizer and checked the performance on the validation set for every epoch. In the GRU, a stopping criterion was used to prevent overfitting. The criterion was that the loss on the validation set needed to be larger than the training set for at least 10 consecutive steps.

In the vARNN model training, backpropagation through a random node is seemingly impossible. We implemented a method called the reparameterization trick, applied in the variational auto-encoders (VAE) for capturing latent features of noisy and high dimensional inputs (Kingma and Welling, 2013). In this method, a separated parameter takes care of the randomness and allows backpropagation to flow through the deterministic nodes.

### 3.6. Model evaluation metrics

We tested whether vAR was a better model for predicting the sequential structure by comparing the residuals of AR, vAR, and GRU. All three models were tested on the simulated and human behavioral data. Specifically, we first trained each model and compared the residual error Res($\theta^*$) on the validation set. In order to compare across subjects, sessions, and models, we have defined the residual as the ratio between the unexplained variance and total data variance. Note that this metric is similar to $1- R^2$, with non-uniform weights on the sample errors. Res = 1 indicates complete structurelessness, and Res = 0 indicates a fully predictable process. We also computed Res from optimization with respect to the randomly weighted error (*rwMSE*) and obtained a null distribution of Res value for the vAR.

Note that the weighted error (*wMSE*) is related to the log-likelihood under the grand assumption of Gaussian distribution. Therefore, minimizing this *wMSE* is equivalent to finding the posterior updated with observation on each trial. We refer to *wMSE* to be consistent with the metrics used in other models. Also note that all models, tested here or reported elsewhere, can only account for approximately 20% of the total variance in the human data, the remaining 80% of the noise is temporally independent (Figure 5A). Other sources of variability can be due to downstream motor noise. Analysis of correlation and reward modulated variability has shown that the noise of saccadic movement and hand pressing was decoupled (Wang et al., 2020), suggesting that the majority likely pertained to the output stage of the motor system.

Figure 4C and 6B show the comparison between models. There were 62 behavioral sessions collected from five human subjects (two females and three males, 18-65 years) who participated in the CSG task. Each behavioral session provided two time series one from eye movement and one from hand movement. This amounted to a total 124 time series and 59,410 trials in the data set. In the model simulation, we repeatedly generated 10 sessions for each parameter set and 1000 trials for each session. For finding the model parameters, we also repeated the optimization 10 times with randomized initialization.

### 4. RESULTS

### 4.1. Behavioral observations

An example session of a human subject performing the button pressing task (Figure 1) is shown in Figure 2A. On average, human subjects received positive feedback ($r = 1$, green) in half of the trials (50.2%). A running mean with the window size of 20 trials has shown that the performance fluctuated around the targeted interval ($e = 0$). Analyzing the partial autocorrelation ($ParCorr$) as the function of trial lag showed that the temporal correlation of the behavioral sequence lasted up to 20 trials (Figure 2B). Analyzing the statistics of the two adjacent trials ($e_{n-1}$ and $e_n$) revealed a nontrivial way the feedback influenced the timing variability (Figure 2C). It's shown that positive reinforcement ($r > 0$) reduced the motor variability while negative ones ($r = 0$) led to increased variability. Additional control experiments that randomly delivered the feedback verified that the feedback was *causally* involved in the adjustment of motor variability (Wang et al., 2020).

Behavioral characterization pointed to a strategy that humans were planning the action by incorporating the past actions and the corresponding feedback in sequential order. A neural network with recurrent units (RNN) provides the generic architecture for this task. We explicitly designed RNNs such that (1) the past performance and feedback were stored in the memory and can be retrieved later; (2) the feedback was used to assess the desirability of corresponding performance and (3) the model used a recurrent unit to generate this process recursively. Based on these requirements, we have tested three types of recurrent units: ARNN, vARNN, and GRU (Figure 3B). All three structures have taken the previous trials as input for the current trial prediction. vARNN and GRU additionally took feedback as input.

**4.2. Model captures the temporal characteristics of human behavior**
We first verified that models could give rise to salient characteristics of observed human behavior. The long-range correlation was recaptured (Figure 3D) by all models in the generated time series. Next, we asked how scalar feedback can be used to improve the estimation of production times, a variable in vector form? One straightforward approach could be that the circuit rewires or redistributes the strengths of connectivity among the memory components. That is to increase the connection ($\omega_i$) associated with the ones that lead to positive feedback and to reduce it for the negative ones. In this thread of thought, GRU was designed to take feedback as input (Figure 3C), and modify the connectivity in a way so that the model would learn from behavioral data. On the other side, vAR was proposed to solve this problem by adjusting the noise level of the weight (Figure 3B inset) instead of adjusting the weight itself.

In human data, the variance increased significantly with the increasing size of the error in the previous trial (Figure 2C). This variance adjustment was most evident in the time series generated with the vAR, but less prominent in the AR and GRU (Figure 3E). This might seem a trivial finding given that the proposed vAR deliberately introduces varying levels of noise, however, as it is demonstrated in the Supplementary materials, vAR actually provides a close approximation to the theoretical optimal given the constraint of the system (Supplementary Figure 1).

**4.2. Model can learn from data**
In the next step, we verified that the vAR model can be trained by the data. We found that the resulting parameters converged to the true parameters reliably (Figure 4A and B, pairwise t-test $p = 2.78e-4$). We also found that the weighted residual error *(wMSE)* provided a better objective function for optimization as it led to a significantly smaller error than the conventional *MSE* (Figure 4C, pairwise t-test, one tail, $p = 3.06e-5$). It also deviated significantly from the null distribution computed with random weights (*rwMSE*) (Figure 4C, p = 6.70e-5 ). In theory, there can be an unbounded number of hidden units in GRU to capture any complex temporal dependencies. We have tested a single GRU layer with varying hidden units, which has shown asymptotic performance. We found the performance of GRU was not comparable to the vAR's (Figure 4C, *p* = 9.02e-4), as the latter has taken into account the uncertainty of error.

## 4.3 Comparing the vARNN model to its alternatives, and comparing model prediction error to its null distribution

Finally, we tested all three models on human behavior. The residual errors of the test set were computed respectively and compared between models after optimization. Across all human behavioral sessions, vAR stood out as the best network model in terms of its accuracy of prediction based on past history (Figure 5A, pairwise t-test, *p* = 1.36e-23 between model AR and vAR, and *p* = 4.47e-29 between GRU and vAR). The network weights $\omega_i$ and variance ratio ($\sigma_+^2/\sigma_-^2$) in vAR were obtained for each behavioral session, and their distributions are shown in Figure 6B. The median value of the variance ratio was 0.226 and significantly smaller (*p* = 1.5e-18) than the null hypothesis where the ratio should approximate 1. These findings confirmed the strategy that a learning agent reduces the amount of noise in response to positive feedback. In simulating the vAR model, we assumed that the two levels of noise were shared among all connections between the input and the AR unit. We think that this is a strong assumption given the observed range of variance across behavioral sessions. Ideally, a connection might have its private and optimized range of noise adjustment. Once vAR was trained for a behavioral session, we made a forward prediction and estimated the variance using the outcomes in the past trials. Figure 6 illustrates the one-step forward estimation of mean and standard error ($\sigma(\hat{e})$). The uncertainty of estimation was dynamically adjusted according to the recent history of feedback (*r*). This additional information is the key feature of our model in comparison to others.

## 5. DISCUSSION

The nervous system is constantly calibrating itself in response to the ever-changing environment using external feedback. The trial-by-trial variations in a motor timing task we observed in humans carried both feedback sensitive and memory related components. This observation is consistent with the idea that animals and humans actively use the motor noise as a means to establish a trade-off between exploitation and exploration., i.e. reducing the noise when the outcome is better than expected (less exploration) and increasing the noise when the outcome becomes worse. Reward-based modulation of motor variability for stochastic exploration during motor learning has been reported or suggested by various previous studies (Dhawale et al., 2019, 2017; Pekny et al., 2015; Wang et al., 2020; Wu et al., 2014). However, the neural circuit mechanisms governing this process are still to be revealed.

In constructing our model, we assumed that the long-term temporal dependency which is unlikely to be adjusted continuously, and therefore, can be modeled by an autoregressive process. This leads to observed long-term stationary temporal correlations observed in our and other paradigms (Chaisanguanthum et al., 2014; Chen et al., 1997; Fischer and Whitney, 2014; Liberman et al., 2014; Murakami et al., 2017). We further assumed that the synaptic connections have fixed values on average, corresponding to the fixed coefficient in the AR model. The short-term adjustment of timing responses was realized by injecting various levels of noise into the synaptic weights in proportion to the strength of the connections leading to the vAR model. The model output provides a probabilistic estimation for the next trial. Note that this model does not apply to a volatile environment where the reward contingency constantly changes, under which circumstance the learning presumably relies on the fast reconfiguration of synaptic strengths. We found that the modeling result is strikingly similar to the optimal control. This reductionist approach allows us to capture the essence of the behavior in a simple recurrent form, and focus on the interpretability of the model.

We evaluated this model and its alternatives according to their prediction accuracy. Training a discriminative neural network model for prediction is practically finding a deterministic mapping between input and output. On the other hand, a generative model estimates the conditional distribution of the observation. Our model uses a hybrid approach. At the input stage, the network takes the previous outcomes as new evidence, and within the recurrent unit, the new evidence is interpreted stochastically by feedback modulated synaptic variability to update the prior. At the output stage, the network computes the prior and likelihood to obtain the posterior, taking advantage of the discriminative features of the point estimation to assimilate the probabilistic estimation. There are other algorithms turning the deterministic regression model into a probabilistic generator. For example, a technique called thinning the network can achieve this by randomly and repeatedly dropping out the nodes or weight updating during training (Srivastava et al., 2014). By randomly turning on and off the nodes, it can significantly reduce overfitting and serve as the regularization. Presumably, it is applied to redundantly connected networks. If we replace each node in the vARNN with a fully connected ensemble of units, our network can effectively implement the sampling algorithm similar to the thinning.

In this task, the only feedback the subjects received is a binary signal about the accuracy of timing at the end of each trial. The dopamine (DA) signaling in the brain provides a candidate infrastructure for such a signal, as elaborated in reinforcement learning (RL) algorithms.DA is involved in the representation of reward and motivation in the brain, and therefore, has been suggested as the neurobiological substrate for the feedback signal in RL. However, unlike that in RL algorithms, in our model, the state of the system is not fully available in order to update the corresponding value function. We here proposed a slightly different mechanism, which requires the brain to adjust its neural noise or, more specifically, the signal-to-noise level. Numerous studies pointed toward the effect of DA on signal and synaptic transmission in the prefrontal cortex (PFC) and PFC-striatum pathways (Law-Tho et al., 1994; Nicola and Malenka, 1997; Tritsch and Sabatini, 2012). This provides the neural substrate for the trial-by-trial adjustment of synaptic noise at the cellular level. Using acute slice preparations, it has been shown that DA can affect the intrinsic membrane excitability and synaptic transmission in the prefrontal cortex (PFC) (Kroener et al., 2009). Appropriate levels of DA receptor (D1R) stimulation can suppress the processing of neural noise and improve animals' performance in a working memory task (Vijayraghavan et al., 2007). It's highly possible that a similar mechanism was at play in our task. The DA projection carrying feedback signal is sent to PFC and the downstream basal ganglia areas. Another relevant example of controlling neuronal noise is attention in perception and cognition. It's been shown that the sensory cortical neurons can vary their correlated spiking noise depending on the relationship between the receptive field and the spatial cues (Ni et al., 2018; Ruff and Cohen, 2014). It's also possible that, by tuning noise among PFC neurons, the brain can harness the correlated variability in the population to achieve optimal behavioral control.

We note that in the behavioral experiments, the motor noise was harnessed differently during different phases of learning. The same neural mechanisms and neurochemical signaling can lead to apparently different strategies implemented at the level of behavior. Intriguingly, we found that in the naive human subjects where a new target was assigned during each test session, their baseline motor variability was maintained at a high level. The positive reinforcement effectively reduces the variability. On the other hand, the well-trained subjects, where the baseline variability was kept at a minimal level, actively increase their motor noise such that exploration leads to fast descent towards the rewarding place. This important observation lends insight into how we should regularize the neural network at different stages of training, and suggests that in the initial phase, the model weight might respond to the positive reinforcement with variance reduction. However, at the expert learning stage where continuous minor adjustment can be accomplished by actively increasing the variance in response to the negative reinforcement.

## 6. CONCLUSION

Overall, we proposed a neural network model that implements the observed behavioral process by feedback-dependent tuning of the noise of connectivity. With its simplicity and biological plausibility, our model is an important supplement to the existing understanding of motor control. Our model also bridges the gap between models of continuous state control in artificial intelligence systems and reinforcement learning in biological systems.

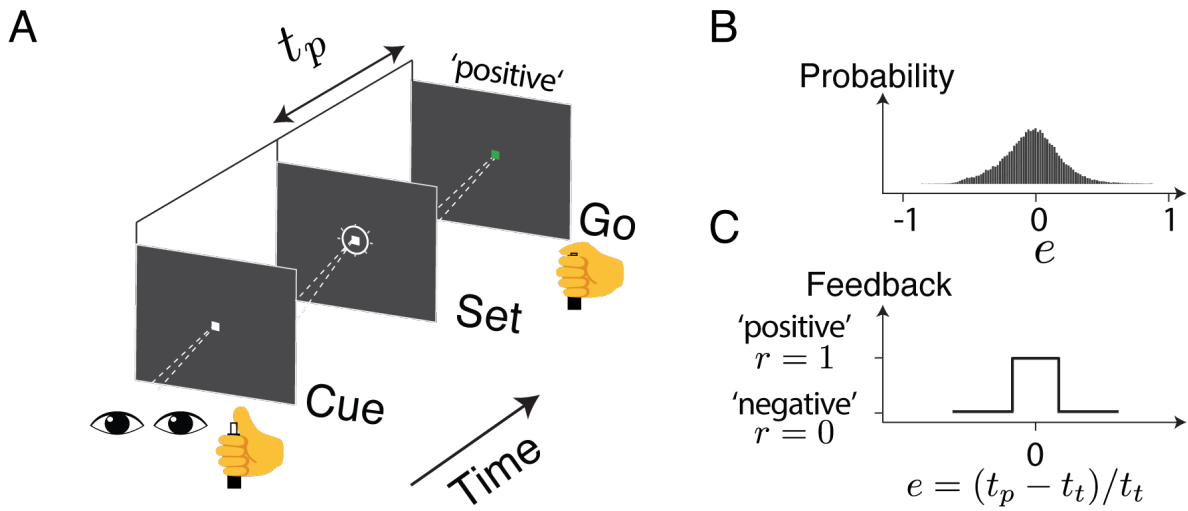

Figure 1. The task design. (A) Human subjects were instructed to make a delayed saccadic eye movement or button press. The produced duration ($t_p$) was compared to a target duration ($t_t$), unbeknownst to the subjects. The relative error is defined as $e = (t_p - t_t)/t_t$). (B) The overall distribution of error across the sessions (C) Based on the timing accuracy, the subject received a visual and auditory cue to indicate either positive or negative feedback.

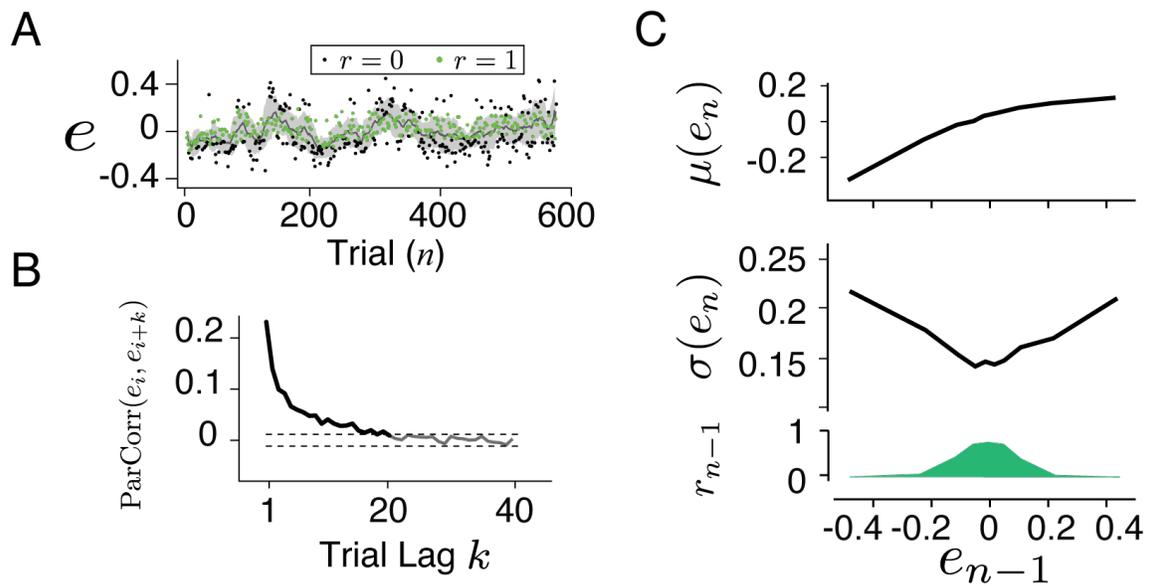

Figure2. Characterizing human behavior. (A) An example time series of the timing error for button presses. The subject was able to adjust his timing according to feedback (green: positive, black: negative) and the trial-by-trial error was fluctuating around the target. (B) The partial autocorrelation coefficient between the errors ($\mathrm{ParCorr}(e_i, e_{i+k})$) as the function of trial lag (k) was calculated. Dash lines are the 1% and 99% confidence levels of the correlation shuffling the trials. The ParCorr value above the 99% level indicated a significantly correlated and the long-term temporal dependency (up to 20 trials) was evident. The first feature which we characterized in the behavior is the slow fluctuation around the target. (C) The mean (top), variance (middle), and the fraction of positive feedback (bottom) were related to the error in the previous trial ($e_{n-1}$). The second feature in the behavior was characterized as the increased variance resulting from the negative feedback.

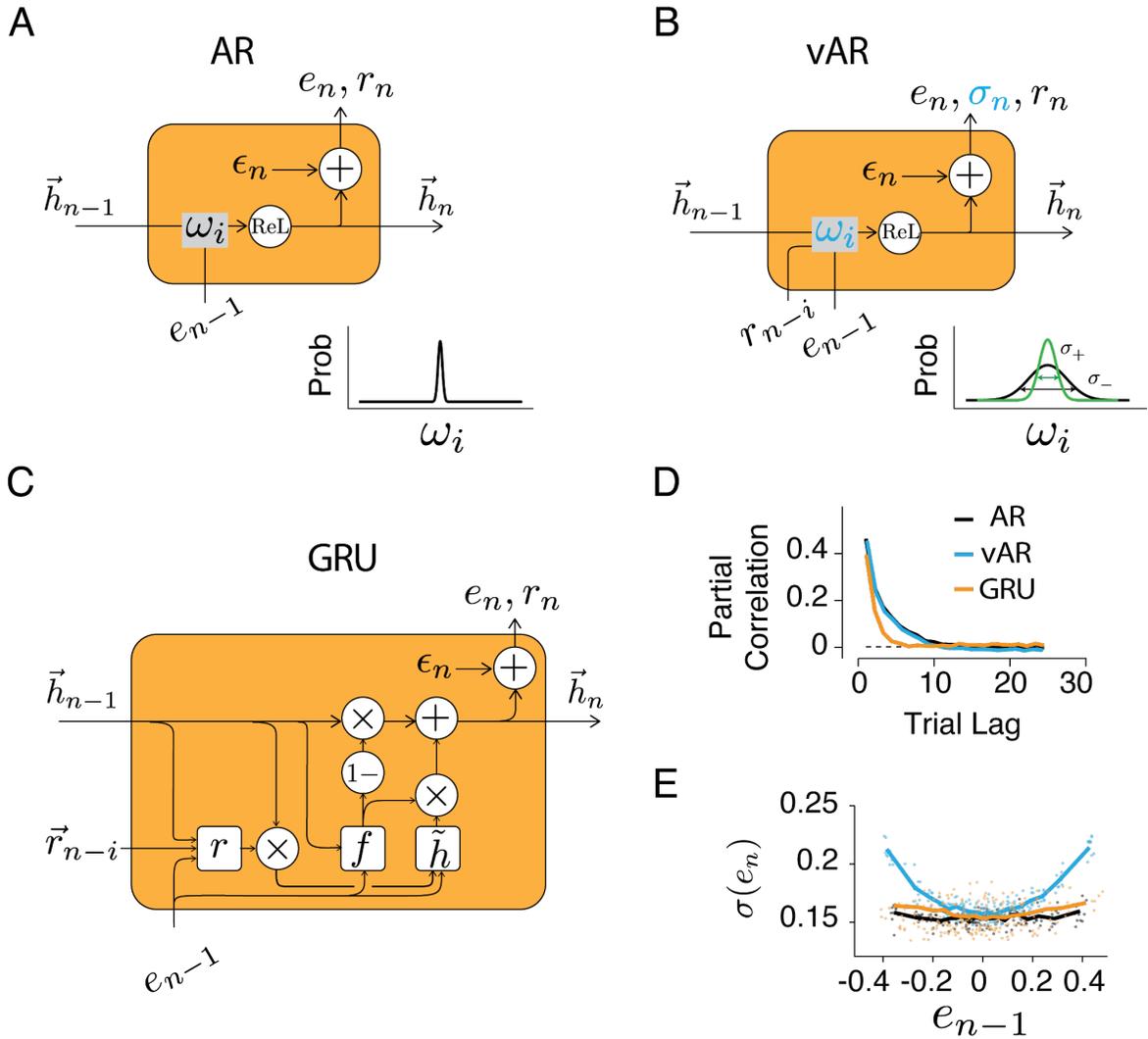

Figure 3. Neural network models. Structure of the recurrent unit in AR (A), vAR (B), and GRU model (C). The output of NN ($e_n$) was retrieved and recursively served as its inputs. In vAR and GRU, the feedback ($r_n$) also served as input. In AR, the input weights were kept constant. In vAR, the input weights were adjusted according to the respective feedback. Specifically, positive feedback reduces the variance of weight (green). The hidden unit ($h_n$) implements the rectified linear function. In GRU, the hidden state was updated based on the computed node values ($r$: reset gate, $f$: forget gate, $\tilde{h}$: candidate state). The private and independent noise ($\epsilon_n$) was sampled from a normal distribution. (D) Time series were simulated for each model and the correlation showed the long-term temporal dependency of the output. (E) The output error variance was modulated by the feedback in vAR similar to the way in human behavior, but not in AR or GRU model.

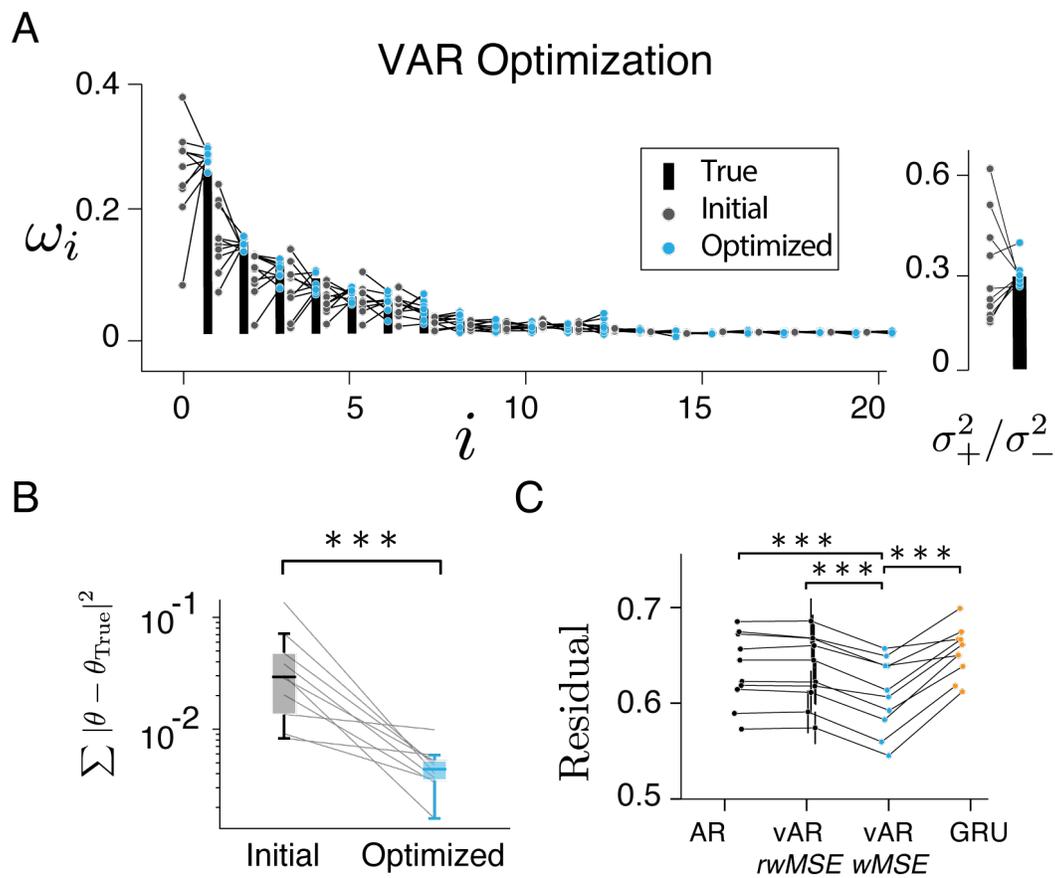

Figure 4. Model performance. (A) and (B) have shown the resulting parameters were reliably converged to their true values. (C) vAR model with weighted MSE (*wMSE*) produced the smallest residual. The residual was normalized to the total variance for easy comparison across sessions and models.

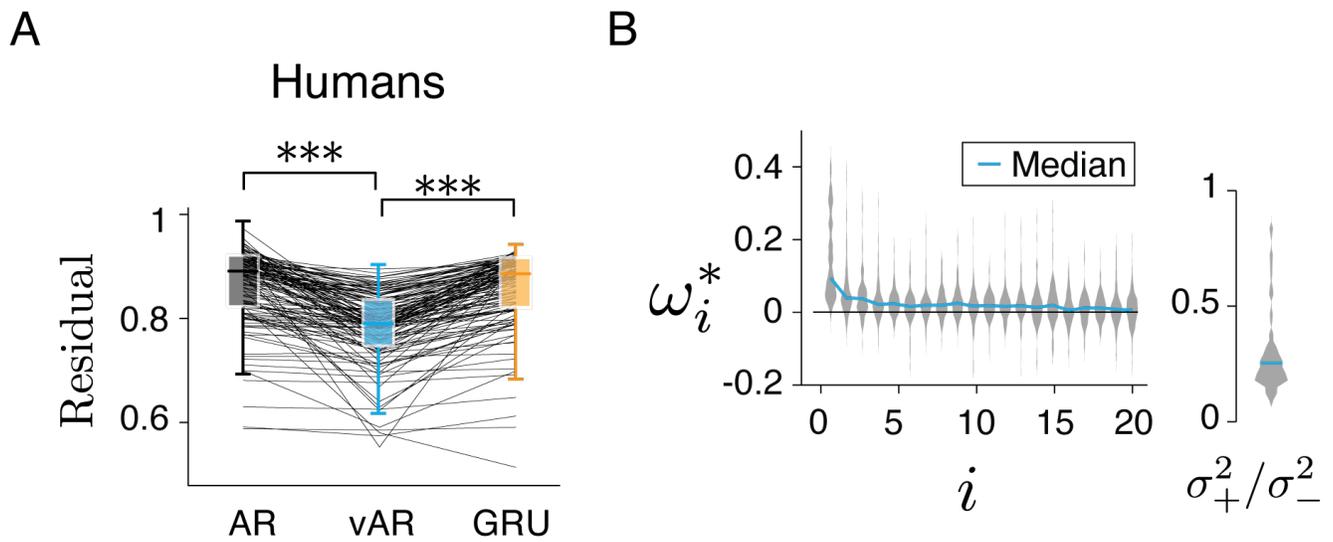

Figure 5. Models and human behavior (A) The vAR was able to capture the data better than the AR and GRU ( *** $p << 0.001$). (B) Similar to figure 4A, vAR parameters were optimized for individual sessions, and the median of the whole distribution was indicated on the violin plot.

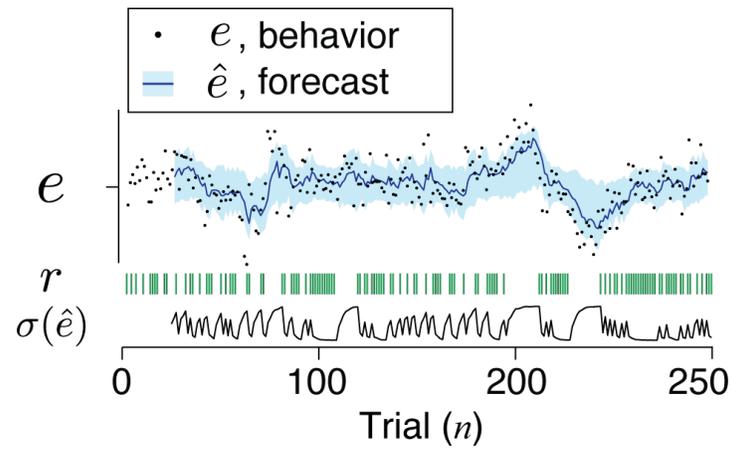

Figure 6. An example sequence of human behavior and its vAR model output. The actual feedback, model forecast, and uncertainty associated with the forecast were shown.

**Data and code availability statement**

Relevant data and code are available at https://github.com/wangjing0/vARNN

**Author Contributions**

JW and IO designed the study. JW developed the code and analysis. YE contributed the code and analysis. All authors contributed to the conceptualization and writing of the manuscript.

**Conflict of Interest**

The research was conducted in the absence of any commercial or financial relationships that could be construed as a potential conflict of interest.

# *Supplementary Materials*

## 1. Supplementary Data and Figures

**Optimal control of the synaptic variance.**

Let's assume that the coefficients have constant means in the vAR, and the system was able to inject noise of various amplitudes to maximize the gain in the next trial. Under those assumptions, the expected error and variance of the error can be expressed as

$$\hat{e}_n = \sum_p \hat{\omega}_i e_{n-i}$$
$$\Sigma^2 = \text{cov}(e_n) = \sum_i \text{cov}(\omega_i) e_{n-i}^2 + n_0^2$$

Hence, the probability was computed by integrating the error distribution and feedback function, shown in Supplementary Figure 1A.

$$\begin{aligned}E(r) &= \frac{1}{\sqrt{2\pi}\Sigma} \int_{-\infty}^{+\infty} r(x)\exp[-\frac{(x-\hat{e})^2}{2\Sigma^2}]dx \\ &= \frac{1}{\sqrt{2\pi}\Sigma} \int_{-D}^{+D} \exp[-\frac{(x-\hat{e})^2}{2\Sigma^2}]dx \\ &= \frac{1}{2}[\text{erf}(\frac{D+\hat{e}}{\sqrt{2}\Sigma}) + \text{erf}(\frac{D-\hat{e}}{\sqrt{2}\Sigma})]\end{aligned}$$

In which $r(x)$ is the square feedback function and D is the acceptance window width. erf(.) is the integral of the Gaussian. To solve the optimization, we have

$$\partial E(r)/\partial \Sigma = -\frac{1}{\sqrt{2\pi}}(\frac{D+\hat{e}}{\Sigma^2}\exp[-\frac{(D+\hat{e})^2}{2\Sigma^2}] + \frac{D-\hat{e}}{\Sigma^2}\exp[-\frac{(D-\hat{e})^2}{2\Sigma^2}])$$

When $D > |\hat{e}|$, $E(r)$ is a monotonically increasing function as $\Sigma$ decreasing. However, $\Sigma$ is bounded and the optimal $\Sigma$ is therefore chosen to be minimal. When $D \leq |\hat{e}|$, we have

$$\partial E(r)/\partial \Sigma = 0 \Rightarrow \Sigma^2 = 2D|\hat{e}|/ln\frac{|D+\hat{e}|}{|D-\hat{e}|}$$

We have the optimal variance ($\Sigma^2_{\text{Analytical}}$) and the model choice of variance ($\Sigma^2_{\text{Approx}}$) as the function of expected error ($\hat{e}$), which is shown in Supplementary Figure 1B. In this illustration, without losing generality, we will assume the first term dominates in the AR relationship, i.e. $\omega_1 >> \omega_2, ...$.

$$\Sigma^2_{\text{Analytical}}(\hat{e}) = \begin{cases} \sigma_0^2 & |\hat{e}| \leq D \\ 2D|\hat{e}|/\ln|\frac{D+\hat{e}}{D-\hat{e}}| & |\hat{e}| > D \end{cases}$$

$$\Sigma^2_{\text{Approx}}(\hat{e}) = \begin{cases} \hat{e}^2\sigma_+^2/\omega_1^2 + \sigma_0^2 & |\hat{e}| \leq D \\ \hat{e}^2\sigma_-^2/\omega_1^2 + \sigma_0^2 & |\hat{e}| > D \end{cases}$$

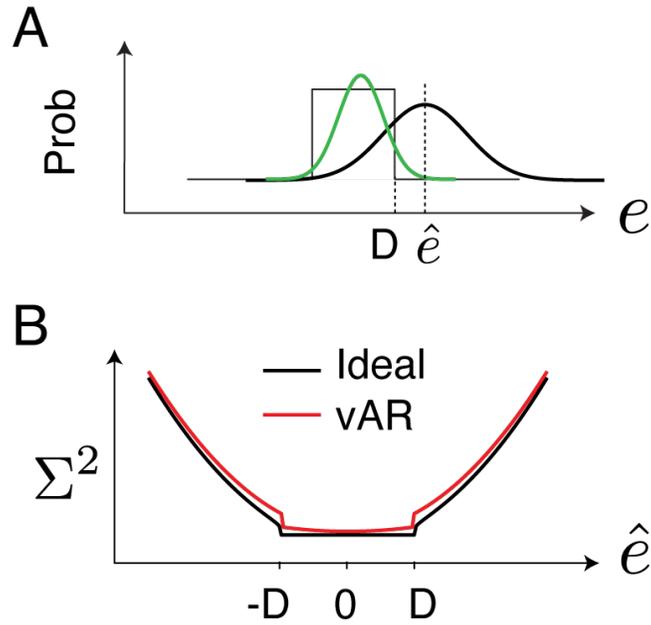

**Supplementary Figure 1**. Optimal variance adjustment. (A) The expected rate of receiving positive feedback is obtained by combining the feedback (Square shape) and the error distribution (Gaussian). In an ideal scenario, the variance of the error could be adjusted freely. It decreases when the estimated error size is small (green) and increases when the error size is large (black). (B) The relationship between variance size and mean of error. For the ideal case, the variance was computed analytically according to the equation. In our model, we have adopted a simple binary choice of variance, and it resulted in a close approximation to the ideal case.